\documentstyle[amssymb,prd,aps]{revtex}
\newtheorem{twierdzenie}{Theorem} \draft
\begin{document}

\title{Theorems on the Renormalization Group Evolution\\ of Quark
Yukawa Couplings and CKM Matrix}

\author{P.~Kielanowski$^{1,2}$, S.R.~Ju\'arez~W.$^3$ and
J.G.~Mora~H.$^2$}

\address{ $^1$~Institute of Theoretical Physics, University of
Bia{\l}ystok, Poland\\ $^2$~Departamento de F\'{\i}sica, CINVESTAV,
M\'{e}xico\\ $^3$~Escuela Superior de F\'{\i }sica y Matem\'{a}ticas, IPN,
M\'{e}xico}

\maketitle
\begin{abstract}
We analyze the two loop renormalization group equations in the
standard model and its extensions for the coupling constants and the
quark Yukawa couplings. The key point of our analysis is the observed
hierarchy of the quark masses and the CKM matrix.  For the one loop
evolution we find the explicit solution for the evolution of the
Yukawa couplings and show the following: 1.~the CKM matrix depends on
the energy on only one function of energy; 2.~the ratios of the down
quark masses depend on the energy through the same function as the CKM
matrix; 3.~the diagonalizing matrices of the biunitary transformation
of the up quarks are energy independent. Next we give the explicit
form of the two loop corrections to the evolution of the quark Yukawa
couplings and show that the relative corrections are of the order \(
\lambda ^{4} \) for the quark Yukawa couplings and of the order \(
\lambda ^{5} \) for the CKM matrix. Finally we give the equations of
the one loop evolution of the squares of the matrix elements of the
CKM matrix and their explicit solution.
\end{abstract}
\pacs{05.10.Cc,11.10.Hi,12.15.Hh}

\section{Introduction}

The problem of the Renormalization Group EVolution (RGEV) of the Quark
Yukawa Couplings (QYC) and the Cabibbo-Kobayashi-Maskawa matrix (CKM)
has been studied in many earlier papers,
Refs.~\cite{ref1,ref2,ref3}. Despite the fact that many, both
analytical end numerical results have been obtained on the evolution
of the QYC and CKM matrix there is no single reference where some
general properties of the evolution have been discussed. This paper is
aimed to fill this gap and it discusses the general properties of
the evolution of the CKM matrix and quark masses that follow from the
hierarchical structure of the QYC.

The information on the properties of the QYC come from the
experimental values of the quark masses and the CKM matrix. Both the
quark masses and the CKM matrix show the hierarchical structure with
the parameter \( \lambda \)\( \approx \sin \theta _{C} \)\( \approx
0.22. \) This hierarchical structure must also be present in the QYC
and is used to derive some conditions for the unknown QYC,
Refs.~\cite{ref4} which reduce the number of parameters of the
standard model.

In this paper we systematically investigate the influence of the
hierarchical structure on the evolution of the CKM matrix and quark
masses by constructing the exact solution of the one loop RGEV
equations compatible with the observed hierarchy and then considering
the two loop RGEV equations with the next order corrections in~\(
\lambda . \)

The most important result that we derive is that the CKM evolution
depends only on one parameter which is a suitable integral that
depends on the model (Standard Model (SM), Minimal Supersymmetric
Standard Model (MSSM) and Double Higgs Model (DHM)). The corrections
to the one loop evolution of the CKM matrix are of the relative
order~\( \lambda ^{5}. \) We next show that the evolution of the
ratios of the down quark masses \( m_{d}/m_{s} \) and \( m_{s}/m_{b}
\) depends on the same parameter as the CKM matrix. The evolution of
the up quark masses is different. Here the quark masses \( m_{u} \),
\( m_{c} \) depend linearly on the corresponding initial values and
their ratio \( m_{u}/m_{c} \) is constant~\cite{ref4a} while the
dependence for \( m_{t} \) is non linear.

The organization of the paper is the following. In Section~II we
introduce the notation and the essential experimental
facts. Afterwards we discuss the RGEV equations up to two loops and
their approximate form compatible with the observed hierarchy. This
gives the basis for the derivation in Section~III of the exact
solutions of the one loop RGEV equations. We also present there the
perturbative scheme for the study of the corrections coming from the
higher order terms in \( \lambda \) and the two loop contributions. In
Section~IV we discuss the properties of the solutions derived earlier
and their physical importance. In Section~V we derive the approximate
evolution equations for the squares of the absolute values of the CKM
matrix elements and give their explicit solution.  In such a way we
give the explicit form of the one loop renormalization group evolution
of the full CKM matrix.  Section~VI is devoted to the conclusions.

\section{Quark Yukawa Couplings and CKM Matrix--Hierarchy and Evolution}

The quark and lepton masses in the standard model arise through the
Higgs mechanism from the Yukawa couplings which have the following
structure in the SM

\begin{equation}
\label{eq1}
\sum ^{3}_{i,j=1}(f^{(e)}_{ij}\overline{e}_{L}^{i}\phi
e^{j}_{R}+(y_{u})_{ij}\overline{u}^{i}_{L}\widetilde{\phi
}u^{j}_{R}+(y_{d})_{ij}\overline{d}^{i}_{L}\phi d^{j}_{R}+\mbox
{h.c.}).
\end{equation}
Here \( f^{(e)} \), \( y_{u} \), and \( y_{d} \) are the matrices of
the Yukawa couplings of leptons and up and down quarks,
respectively. \( \phi \)~is the scalar Higgs field. The theory itself
does not put any restrictions or conditions on the \( f^{(e)} \), \(
y_{u} \) and \( y_{d} \) matrices and from the phenomenological point
of view they are constrained by the values of the physical lepton and
quark masses

\begin{equation}
\label{eq2}
\mbox {Diag}(m_{u},m_{c},m_{t})=(U_{u})^{\phantom {\dagger
}}_{L}y_{u}(U_{u})^{\dagger }_{R},\qquad \mbox
{Diag}(m_{d},m_{s},m_{b})=(U_{d})^{\phantom {\dagger
}}_{L}y_{d}(U_{d})^{\dagger }_{R}
\end{equation}
with diagonal elements being the up and down quark masses after the
spontaneous symmetry breaking and the CKM matrix. The diagonalizing
matrices \( (U_{u,d})^{\phantom {\dagger }}_{L,R} \) of the biunitary
transformations transform the quark fields in Eq.~(\ref{eq1}) into the
physical quark fields. As a consequence the unitary matrix (CKM
matrix)
\begin{equation}
\label{eq3}
V_{CKM}=(U_{u})^{\phantom {\dagger }}_{L}(U_{d})^{\dagger }_{L}
\end{equation}
 appears in the charged current.

The observed hierarchy of the quark and lepton masses and the CKM
matrix is the following, Refs.~\cite{ref5,ref6}
\begin{equation}
\label{eq4}
\frac{m_{u}}{m_{c}}\sim \lambda ^{4},\quad \frac{m_{c}}{m_{t}}\sim
\lambda ^{4},\quad \frac{m_{d}}{m_{s}}\sim \lambda ^{2},\quad
\frac{m_{s}}{m_{b}}\sim \lambda ^{2},\quad \frac{m_{b}}{m_{t}}\sim
\lambda ^{2},\quad \frac{m_{e}}{m_{\mu }}\sim \lambda ^{4},\quad
\frac{m_{\mu }}{m_{\tau }}\sim \lambda ^{2},\quad \frac{m_{\tau
}}{m_{t}}\sim \lambda ^{3},
\end{equation}
\begin{equation}
\label{eq5}
\left( \begin{array}{ccc} 1-\frac{1}{2}\lambda ^{2} & \lambda &
\lambda ^{3}A(\rho -i\eta )\\ -\lambda & 1-\frac{1}{2}\lambda ^{2} &
\lambda ^{2}A\\ \lambda ^{3}A(1-\rho -i\eta ) & -\lambda ^{2}A & 1
\end{array}\right) 
\end{equation}
with \( \lambda \approx 0.22. \)

The hypothesis of grand unification assumes that the 3 coupling
constants of the standard model converge at the scale \( 10^{15-16}
\)~GeV and the symmetry group of the model becomes larger
(e.g. SU(5)). This may also be an origin of the additional symmetries
or textures of the QYC at this energy scale. To relate the parameters
of the QYC at the GU scale with the observables at low energy one
conventionally uses the Renormalization Group Equations (RGE) for the
coupling constants and the quark and lepton Yukawa couplings. The
structure of the two loop RGE is the following
Refs.~\cite{ref1,ref2,ref3,ref7}
\begin{mathletters}
\begin{equation}
\label{eq6}
\frac{dg_{l}}{dt}=\frac{1}{(4\pi )^{2}}b_{l}g^{3}_{l}-\frac{1}{(4\pi
)^{4}}G_{l}g^{3}_{l},
\end{equation}
\begin{equation}
\label{eq7}
\frac{dy_{u,d,e,\nu}}{dt}=\left[ \frac{1}{(4\pi )^{2}}\beta
_{u,d,e,\nu}^{(1)}+\frac{1}{(4\pi )^{4}}\beta
_{u,d,e,\nu}^{(2)}\right] y_{u,d,e,\nu}.
\end{equation}
\label{eq6s}
\end{mathletters}
The variable \( t \) is defined as \( t=\ln (E/\mu ) \) and the 
constants \( b_{l} \) and functions \( G_{l} \), \( \beta
_{u,d,e,\nu}^{(1)} \) and \( \beta _{u,d,e,\nu}^{(2)} \) are defined
for various models in Appendix~I.  The coefficients \( G_{l}, \) \(
\beta _{u,d,e,\nu}^{(1)}, \) \( \beta _{u,d,e,\nu}^{(2)} \) are
functions of \( g_{l} \) and \( y_{u,d,e,\nu} \) so Eqs.~(\ref{eq6s})
form a system of coupled non linear equations and their explicit
solution is not known. The Yukawa couplings \( y_{u,d,e,\nu} \) are
normalized in the following way
\begin{equation}
\label{eq8}
[y_{u}]_{33}\sim 1,\quad [y_{d}]_{33}\sim \frac{m_{b}}{m_{t}}\sim
\lambda ^{2},\quad [y_{e}]_{33}\sim \frac{m_{\tau }}{m_{t}}\sim
\lambda ^{3}.
\end{equation}
The normalization in Eq.~(\ref{eq8}) is the origin of the hierarchy in
Eqs.~(\ref{eq6s}) because the functions \( G_{l}, \) \( \beta
_{u,d,e,\nu}^{(1)} \) and \( \beta _{u,d,e,\nu}^{(2)} \) contain the
squares and higher powers of \( y_{u,d,e,\nu}. \)

From Eq.~(\ref{eq8}) it follows that positive powers of \( y_{d,e,\nu}
\) are much smaller than the corresponding powers of \( y_{u} \). This
information enables to find the solution of Eqs.~(\ref{eq6s}) and to
study the properties of the solutions. The other origin of the
hierarchy in Eqs.~(\ref{eq6s}) is related to the number of loops. Here
the inclusion of each additional loop is suppressed by the factor \(
1/(4\pi )^{2}\sim \lambda ^{4} \) which is of the same order as \(
([y_{d}]_{33})^{2} \) and the correct approximation procedure must
take this into account.

Eqs.~(\ref{eq6s}) will be solved and analyzed in the following steps:

\begin{enumerate}
\item Definition of the hierarchy of the equations;
\item \emph{Exact} solution of the basic equations;
\item Perturbative corrections to the exact solution;
\item Implications of the former results for the quark masses and the
CKM matrix
\end{enumerate}
In the rest of the paper we will discuss each step in detail.

Eqs.~(\ref{eq6s}) are the two loop RGE for coupling constants \( g_{l}
\) and Yukawa couplings \( y_{u,d,e,\nu}. \) In these equations the
terms of different order in \( \lambda \) are present. As the first
step we will find the approximate form of these equations neglecting
all the terms of \( \lambda ^{4} \) and higher. These approximate
equations have the following form
\begin{mathletters}
\begin{equation}
\label{eq9}
\frac{dg_{l}}{dt}=\frac{1}{(4\pi )^{2}}b_{l}g^{3}_{l},
\end{equation}
\begin{equation}
\label{eq10}
\frac{dy_{u}}{dt}=\frac{1}{(4\pi )^{2}}[\alpha ^{u}_{1}(t)+\alpha
^{u}_{2}y^{\phantom {\dagger }}_{u}y^{\dagger }_{u}+\alpha
^{u}_{3}\mbox {Tr}(y^{\phantom {\dagger }}_{u}y^{\dagger
}_{u})]y_{u}=\frac{1}{(4\pi )^{2}}E^{u}_{1}y_{u},
\end{equation}
\begin{equation}
\label{eq11}
\frac{dy_{d}}{dt}=\frac{1}{(4\pi )^{2}}[\alpha ^{d}_{1}(t)+\alpha
^{d}_{2}y^{\phantom {\dagger }}_{u}y^{\dagger }_{u}+\alpha
^{d}_{3}\mbox {Tr}(y^{\phantom {\dagger }}_{u}y^{\dagger
}_{u})]y_{d}=\frac{1}{(4\pi )^{2}}E^{d}_{1}y_{d},
\end{equation}
\begin{equation}
\label{eq12}
\frac{dy_{e}}{dt}=\frac{1}{(4\pi )^{2}}[\alpha ^{e}_{1}(t)+\alpha
^{e}_{2}y^{\phantom {\dagger }}_{u}y^{\dagger }_{u}+\alpha
^{e}_{3}\mbox {Tr}(y^{\phantom {\dagger }}_{u}y^{\dagger
}_{u})]y_{e}=\frac{1}{(4\pi )^{2}}E^{e}_{1}y_{e},
\end{equation}
\begin{equation}
\label{eq12a}
\frac{dy_{\nu}}{dt}=\frac{1}{(4\pi )^{2}}[\alpha ^{\nu}_{1}(t)+\alpha
^{\nu}_{2}y^{\phantom {\dagger }}_{u}y^{\dagger }_{u}+\alpha
^{\nu}_{3}\mbox {Tr}(y^{\phantom {\dagger }}_{u}y^{\dagger
}_{u})]y_{\nu}=\frac{1}{(4\pi )^{2}}E^{\nu}_{1}y_{\nu}
\end{equation}
\label{eq8s}
\end{mathletters}
where the values of \( \alpha ^{u,d,e,\nu}_{i} \)
are given in Appendix. Eqs.~(\ref{eq8s}) will be solved exactly and
as a next step we will consider Eqs.~(\ref{eq6s}) where the terms of
the order \( \lambda ^{4} \) are kept:
\begin{mathletters}
\begin{equation}
\label{eq13}
\frac{dg_{l}}{dt}=\frac{1}{(4\pi
)^{2}}b_{l}g^{3}_{l}-\frac{g^{3}_{l}}{(4\pi )^{4}}\left(
C_{lu}\mbox {Tr}(y_{u}y^{\dagger }_{u}) + \sum
_{k}b_{kl}g^{2}_{k}\right),
\end{equation}
\begin{eqnarray}
\frac{dy_{u}}{dt} & = & \frac{1}{(4\pi
 )^{2}}E^{u}_{1}y_{u}+\frac{1}{(4\pi )^{2}}[\alpha
 ^{u}_{4}y_{d}y^{\dagger }_{d}+\alpha ^{u}_{5}\mbox
 {Tr}(y_{d}y^{\dagger }_{d})]y_{u}\nonumber \\ & + & \frac{1}{(4\pi
 )^{4}}F^{u}(y_{u}y^{\dagger }_{u},g_{l})y_{u}=\frac{1}{(4\pi
 )^{2}}E^{u}_{1}y_{u}+\frac{1}{(4\pi )^{4}}E^{u}_{2}y_{u},\label{eq14}
\end{eqnarray}
\begin{eqnarray}
\frac{dy_{d}}{dt} & = & \frac{1}{(4\pi
 )^{2}}E^{d}_{1}y_{d}+\frac{1}{(4\pi )^{2}}[\alpha
 ^{d}_{4}y_{d}y^{\dagger }_{d}+\alpha ^{d}_{5}\mbox
 {Tr}(y_{d}y^{\dagger }_{d})]y_{d}\nonumber \\ & + & \frac{1}{(4\pi
 )^{4}}F^{d}(y_{u}y^{\dagger }_{u},g_{l})y_{d}=\frac{1}{(4\pi
 )^{2}}E^{d}_{1}y_{d}+\frac{1}{(4\pi )^{4}}E^{d}_{2}y_{d},\label{eq15}
\end{eqnarray}
\begin{eqnarray}
\frac{dy_{e}}{dt} & = & \frac{1}{(4\pi
 )^{2}}E^{e}_{1}y_{e}+\frac{1}{(4\pi )^{2}}[\alpha
 ^{e}_{4}y_{d}y^{\dagger }_{d}+\alpha ^{e}_{5}\mbox
 {Tr}(y_{d}y^{\dagger }_{d})]y_{e}\nonumber \\ & + & \frac{1}{(4\pi
 )^{4}}F^{e}(y_{u}y^{\dagger }_{u},g_{l})y_{e}=\frac{1}{(4\pi
 )^{2}}E^{e}_{1}y_{e}+\frac{1}{(4\pi )^{4}}E^{e}_{2}y_{e},\label{eq16}
\end{eqnarray}
\begin{eqnarray}
\frac{dy_{\nu}}{dt} & = & \frac{1}{(4\pi
 )^{2}}E^{\nu}_{1}y_{\nu}+\frac{1}{(4\pi )^{2}}[\alpha
 ^{\nu}_{4}y_{d}y^{\dagger }_{d}+\alpha ^{\nu}_{5}\mbox
 {Tr}(y_{d}y^{\dagger }_{d})]y_{\nu}\nonumber \\ & + & \frac{1}{(4\pi
 )^{4}}F^{\nu}(y_{u}y^{\dagger }_{u},g_{l})y_{\nu}=\frac{1}{(4\pi
 )^{2}}E^{\nu}_{1}y_{\nu}+\frac{1}{(4\pi
 )^{4}}E^{\nu}_{2}y_{\nu}.\label{eq16a}
\end{eqnarray}
\label{eq9s}
\end{mathletters}
The functions \(
 F^{u,d,e,\nu}(y_{u}y^{\dagger }_{u},g_{l}) \) are obtained from the
functions \( \beta _{u,d,e,\nu}^{(2)} \) given in Appendix by
putting \( y_{d}=y_{e}=0. \) Eqs.~(\ref{eq9s}) cannot be explicitly
solved but they allow the perturbative solution by transforming them
into integral equations. This allows the study of the corrections to
the exact solutions of Eqs.~(\ref{eq8s}).

\section{Solution of the Renormalization Group Equations for
\protect\( \lowercase {y_{u}}\protect \) and \protect\( \lowercase
{y_{d}}\protect \)}

The renormalization group equations (\ref{eq6s}) will be solved in two
steps.  We will start by solving the one loop
equations~(\ref{eq8s}). Eqs.~(\ref{eq9}) are very easy to solve and
their solution reads
\begin{equation}
\label{eq17}
g_{l}(t)=\frac{g_{l}(t_{0})}{\sqrt{1-
\frac{2b_{l}g^{2}_{l}(t_{0})(t-t_{0})}{(4\pi)^{2}}}}
\end{equation}
Eq.~(\ref{eq10}) is decoupled from Eqs.~(\ref{eq11}) and~(\ref{eq12})
and is non linear. Eqs.~(\ref{eq11}) and~(\ref{eq12}) become linear
once \( y_{u} \)-- the solution of Eq.~(\ref{eq10}) is known. We
therefore solve Eq.~(\ref{eq10}) first.

\subsection{The evolution of \protect\( y_{u}\protect \) from one loop RGE}

The crucial fact enabling the exact solution of Eq.~(\ref{eq10}) is
the observation that the diagonalizing matrices of the biunitary
transformation do not depend on the energy. As mentioned before \(
y_{u} \) can be diagonalized by the biunitary transformation,
Eq.~(\ref{eq2}). The unitary matrices \( (U_{u})_{L,R} \) are the
diagonalizing matrices of the hermitian matrices \(
H_{u}^{1}=y^{\phantom {\dagger }}_{u}(t)y^{\dagger }_{u}(t) \) and \(
H_{u}^{2}=y^{\dagger }_{u}(t)y^{\phantom {\dagger }}_{u}(t) \) and
they fulfill the equations which follow from Eq.~(\ref{eq10})
\begin{equation}
\label{eq18}
\frac{d}{dt}H_{u}^{i}(t)=
\frac{2}{(4\pi)^2}\{\alpha ^{u}_{1}(t)+\alpha
^{u}_{2}H_{u}^{i}+\alpha ^{u}_{3}\mbox
{Tr}(H_{u}^{i})\}H_{u}^{i},\qquad i=1,2.
\end{equation}
The solution of Eq.~(\ref{eq18}) can be written in the following form
\begin{equation}
\label{eq19}
H_{u}^{i}(t)=H_{u}^{i}(t_{0})+\sum ^{\infty
}_{k=1}\frac{(t-t_{0})^{k}}{k!}\left. \frac{d^{k}H_{u}^{i}(t)}{dt^{k}}\right|
_{t=t_{0}}.
\end{equation}
From Eq.~(\ref{eq18}) one can see that the derivative \(
\left. \frac{d^{k}H_{u}^{i}}{dt^{k}}\right| _{t=t_{0}} \) is the sum
of the powers of \( H_{u}^{i}(t_{0}) \) with scalar coefficients.
Thus from the hermiticity of \( H_{u}^{i}(t_{0}) \) it follows that \(
H_{u}^{i}(t) \) are diagonalized by the same matrices as \(
H_{u}^{i}(t_{0}) \) so the diagonalizing matrices \( (U_{u})_{L} \)
and \( (U_{u})_{R} \) of \( H_{u}^{1}(t) \) and \( H_{u}^{2}(t) \) are
energy independent. Thus the matrix \( y_{u}(t) \) has the following
representation
\begin{equation}
\label{eq20}
y_{u}(t)=(U_{u})^{\dagger }_{L}\Delta _{u}(t)(U_{u})^{\phantom
{\dagger }}_{R}
\end{equation}
where on the right hand side of Eq.~(\ref{eq20}) only the diagonal
matrix \( \Delta_{u} (t)=\mbox {Diag}(m_{u}(t),m_{c}(t),m_{t}(t)) \)
does depend on~\( t \) and~\( (U_{u})_{L} \) and~\( (U_{u})_{R} \) are
constant matrices that diagonalize~\( y_{u}(t_{0}) \).
The matrix~\( \Delta _{u}(t) \) fulfills the following equation
\begin{equation}
\label{eq21}
\frac{d}{dt}\Delta _{u}=\frac{1}{(4\pi)^2}
\{\alpha ^{u}_{1}(t)+\alpha ^{u}_{2}\Delta
^{2}_{u}+\alpha ^{u}_{3}\mbox {Tr}(\Delta ^{2}_{u})\}\Delta _{u}.
\end{equation}
Eq.~(\ref{eq21}) splits into 3 equations for the up quark
masses
\begin{mathletters}
\begin{eqnarray}
\frac{dm_{u}}{dt} & = & \frac{1}{(4\pi)^2}
\{\alpha ^{u}_{1}(t)+\alpha
^{u}_{3}m^{2}_{t}\}m_{u},\label{eq22} \\
 \frac{dm_{c}}{dt} & = & \frac{1}{(4\pi)^2}
\{\alpha ^{u}_{1}(t)+\alpha ^{u}_{3}m^{2}_{t}\}m_{c},\label{eq23} \\
\frac{dm_{t}}{dt} & = & \frac{1}{(4\pi)^2}
\{\alpha ^{u}_{1}(t)+(\alpha ^{u}_{2}+\alpha
^{u}_{3})m^{2}_{t}\}m_{t}.\label{eq24}
\end{eqnarray}
\end{mathletters}
We first solve Eq.~(\ref{eq24})
\begin{equation}
\label{eq25}
m_{t}(t)=\frac{m_{t}(t_{0})r^{1/2}_{g}(t)}{\sqrt{1-\frac{2}{(4\pi)^2}
(\alpha^{u}_{2}+\alpha ^{u}_{3})m^{2}_{t}(t_{0})
\int ^{t}_{t_{0}}r_{g}(\tau)d\tau }}
\end{equation}
where
\begin{equation}
\label{eq26}
r_{g}(t)=\exp (\frac{2}{(4\pi)^2}
\int ^{t}_{t_{0}}\alpha ^{u}_{1}(\tau )d\tau ).
\end{equation}
Eqs.~(\ref{eq22}) and~(\ref{eq23}) are identical and their solution
reads
\begin{equation}
\label{eq27}
m_{u,c}(t)=m_{u,c}(t_{0})r^{1/2}_{g}(t)\exp (\frac{1}{(4\pi)^2}
\alpha_{2}^{u}\int
^{t}_{t_{0}}m^{2}_{t}(\tau )d\tau ).
\end{equation}
The full energy evolution of \( y_{u}(t) \) is determined from
Eq.~(\ref{eq20}) using Eqs.~(\ref{eq25}) and~(\ref{eq27}).

\subsection{The evolution of \protect\( y_{d}\protect \) from one loop RGE}

The one loop evolution of \( y_{d} \) is determined from
Eq.~(\ref{eq11}).  The fact that the diagonalizing matrices of \(
y_{u}(t) \) do not depend on \( t \) simplifies Eq.~(\ref{eq11}) and
justifies the following substitution
\begin{equation}
\label{eq28}
y_{d}(t)=(U_{u})^{\dagger }_{L}W(t)
\end{equation}
so \( W(t) \) fulfills the equation
\begin{equation}
\label{eq29}
\frac{dW}{dt}=\frac{1}{(4\pi)^2}
\{\alpha ^{d}_{1}(t)+\alpha ^{d}_{2}\Delta
^{2}_{u}+\alpha ^{d}_{3}\mbox {Tr}(\Delta ^{2}_{u})\}W.
\end{equation}
The matrix on the right hand side of Eq.~(\ref{eq29}) has the
following form
\begin{eqnarray}
 & & \{\alpha ^{d}_{1}(t)+\alpha ^{d}_{2}\Delta ^{2}_{u}+\alpha
 ^{d}_{3}\mbox {Tr}(\Delta ^{2}_{u})\}\nonumber \\ & & =(\alpha
 ^{d}_{1}(t)+\alpha ^{d}_{3}m^{2}_{t})\left( \begin{array}{ccc} 1 & 0
 & 0\\ 0 & 1 & 0\\ 0 & 0 & 1
\end{array}\right) +\alpha ^{d}_{2}m^{2}_{t}\left( \begin{array}{ccc}
0 & 0 & 0\\ 0 & 0 & 0\\ 0 & 0 & 1
\end{array}\right) \label{eq30} 
\end{eqnarray}
and the solution \( W(t) \) of Eq.~(\ref{eq29}) is the following
\begin{equation}
\label{eq31}
W(t)=(r^{\prime }_{g}(t))^{1/2}\exp (\frac{1}{(4\pi)^2}
\alpha ^{d}_{3}\int
^{t}_{t_{0}}m^{2}_{t}(\tau )d\tau )\cdot Z(t)\cdot W(t_{0})
\end{equation}
where
\begin{equation}
\label{eq32}
r^{\prime }_{g}(t)=\exp (\frac{2}{(4\pi)^2}
\int ^{t}_{t_{0}}\alpha ^{d}_{1}(\tau )d\tau
)
\end{equation}
and
\begin{equation}
\label{eq33}
Z(t)=\left( \begin{array}{ccc} 1 & 0 & 0\\ 0 & 1 & 0\\ 0 & 0 & h(t)
\end{array}\right) 
\end{equation}
with
\begin{equation}
\label{eq34}
h(t)=\exp (\frac{1}{(4\pi)^2}
\alpha ^{d}_{2}\int ^{t}_{t_{0}}m^{2}_{t}(\tau )d\tau ).
\end{equation}
Putting Eqs.~(\ref{eq28})--(\ref{eq34}) together we obtain the
following result for the one loop evolution of \( y_{d}(t) \)
\begin{equation}
\label{eq35}
y_{d}(t)=(r^{\prime }_{g}(t))^{1/2}(h(t))^{(\alpha ^{d}_{3}/\alpha
^{d}_{2})}(U_{u})^{\dagger }_{L}Z(t)(U_{u})^{\phantom {\dagger
}}_{L}y_{d}(t_{0}).
\end{equation}

\subsection{Higher order corrections to the \protect\( y_{u}\protect \) and \protect\( y_{d}\protect \)
evolution}

The two loop RGE for \( y_{u} \) and \( y_{d} \), Eqs.~(\ref{eq14})
and~(\ref{eq15}) have the following structure
\begin{equation}
\label{eq36}
\frac{dy_{u,d}}{dt}=\frac{1}{(4\pi
)^{2}}E^{u,d}_{1}y_{u,d}+\frac{1}{(4\pi )^{4}}E^{u,d}_{2}y_{u,d}.
\end{equation}
In Eqs.~(\ref{eq36}) the terms containing the functions \( E^{u,d}_{2}
\) are suppressed by the factor \( 1/(4\pi )^{2}\sim \lambda ^{4} \)
in comparison to the terms with \( E^{u,d}_{1}. \) Eqs.~(\ref{eq36})
can be transformed into the integral equations
\begin{equation}
\label{eq37}
y_{u,d}(t)=y_{u,d}(t_{0})+\frac{1}{(4\pi )^{2}}\int
^{t}_{t_{0}}E^{u,d}_{1}y_{u,d}(\tau )d\tau +\frac{1}{(4\pi )^{4}}\int
^{t}_{t_{0}}E^{u,d}_{2}y_{u,d}(\tau )d\tau
\end{equation}
and Eqs.~(\ref{eq37}) can be transformed into the following recurrence
relation
\begin{equation}
\label{eq38}
y^{(n)}_{u,d}(t)=y_{u,d}(t_{0})+\frac{1}{(4\pi )^{2}}\int
^{t}_{t_{0}}E^{u,d}_{1}y^{(n-1)}_{u,d}(\tau )d\tau +\frac{1}{(4\pi
)^{4}}\int ^{t}_{t_{0}}E^{u,d}_{2}y^{(n-1)}_{u,d}(\tau )d\tau .
\end{equation}
The functions \( y^{(n)}_{u,d}(t) \) converge for \( n\rightarrow
\infty \) to the solution of Eq.~(\ref{eq36}) for an arbitrary initial
function \( y^{(0)}_{u,d}(t) \).  If we choose \( y^{(0)}_{u,d}(t) \)
to be the solution of the one loop RGE, Eqs.~(\ref{eq20})
and~(\ref{eq35}) then we obtain
\begin{equation}
\label{eq39}
y^{(1)}_{u,d}(t)=y_{u,d}(t_{0})+\frac{1}{(4\pi )^{2}}\int
^{t}_{t_{0}}E^{u,d}_{1}y^{(0)}_{u,d}d\tau +\frac{1}{(4\pi )^{4}}\int
^{t}_{t_{0}}E^{u,d}_{2}y^{(0)}_{u,d}d\tau .
\end{equation}
Now
\begin{equation}
\label{eq40}
y_{u,d}(t_{0})+\frac{1}{(4\pi )^{2}}\int
^{t}_{t_{0}}E^{u,d}_{1}y^{(0)}_{u,d}d\tau =y^{(0)}_{u,d}(t)
\end{equation}
because \( y^{(0)}_{u} \) and \( y^{(0)}_{d} \) fulfill
Eqs.~(\ref{eq10}) and~(\ref{eq11}), respectively. It thus follows that
\begin{equation}
\label{eq41}
y^{(1)}_{u,d}(t)=y^{(0)}_{u,d}(t)+\frac{1}{(4\pi )^{4}}\int
^{t}_{t_{0}}E^{u,d}_{2}y^{(0)}_{u,d}d\tau
\end{equation}
which means that the lowest order corrections to the solution of the
one loop RGE for \( y^{(0)}_{u,d}(t) \) are of the relative order \(
\lambda ^{4} \) when compared to the \( y^{(0)}_{u,d}(t) \) and
Eq.~(\ref{eq41}) gives the explicit form of this correction.

\section{Properties of the Renormalization Group Evolution of the Quark Masses and the
CKM Matrix}

In this section we will study the physical implications of the
results obtained in the previous section. We will present them in the
series of the theorems
\begin{twierdzenie}The one loop RGEV of the CKM matrix depends on only one function
of the energy~\( h(t) \), given in Eq.~(\ref{eq34}).\end{twierdzenie}
\emph{Proof:} The CKM matrix, given in Eq.~(\ref{eq3}) is constructed
from the matrices \( (U_{u})^{\phantom {\dagger }}_{L} \) and \(
(U_{d})^{\phantom {\dagger }}_{L} \). In section~III.A we have shown
that \( (U_{u})^{\phantom {\dagger }}_{L} \) does not depend on the
energy.  The dependence on~\( t \) can only be contained in \(
(U_{d})^{\phantom {\dagger }}_{L} \) which is determined from the
matrix \( y_{d}(t) \) of the down QYC. The matrices \(
(U_{d})^{\phantom {\dagger }}_{L} \) and \( (U_{d})^{\phantom {\dagger
}}_{R} \) do not depend on the normalization of \( y_{d}(t) \). From
Eq.~(\ref{eq35}) one can see that the only dependence of \( y_{d}(t)
\) on \( t \) apart from the normalization is contained in the matrix
\( Z(t) \), which is the diagonal matrix and as can be seen from
Eq.~(\ref{eq33}) it depends only on the function \( h(t). \) This
completes the proof of Theorem~1.\begin{twierdzenie}The ratios of the
down quark masses \( m_{d}/m_{s} \) and \( m_{s}/m_{b} \) are the
functions of only \( h(t) \), given in
Eq.~(\ref{eq34})\end{twierdzenie}\emph{Proof:} The down quark masses
are obtained after the diagonalization of the down QYC \( y_{d}(t) \)
and the ratios of \( m_{d}/m_{s} \) and \( m_{s}/m_{b} \) do not
depend on the normalization of \( y_{d}(t). \) It thus follows from
the proof of Theorem~1 that these ratios are the functions of \( h(t)
\).\begin{twierdzenie}The ratio of the up quark masses \( m_{u}/m_{c}
\) is energy independent.\end{twierdzenie}\emph{Proof:} This theorem
follows directly from Eq.~(\ref{eq27}).\begin{twierdzenie}The next
order corrections\footnote{
we include in RGEV equations the terms of the order \( \lambda ^{4} \)
that come from the one loop RGE that contain the terms \( y^{\phantom
{\dagger }}_{d}y^{\dagger }_{d} \) and the two loop contributions of
the order \( \sim 1 \) multiplied by \( 1/(4\pi )^{4}. \) }of the RGEV
of the CKM matrix are of the order \( \lambda
^{5}. \)\end{twierdzenie}\emph{Proof:} If we consider RGE in the next
order then the terms of the order \( \lambda ^{4} \) are preserved in
the equations and the evolution is governed by Eqs.~(\ref{eq9s}) and
the explicit solution in the next leading order is given by
Eq.~(\ref{eq41}).  From Eq.~(\ref{eq41}) one obtains by the direct
calculation the following result for the
commutators
\begin{mathletters}
\begin{eqnarray}
 & & [y_{u,d}^{(1)}y_{u,d}^{(1)^{\dagger
 }},y_{u,d}^{(0)}y_{u,d}^{(0)^{\dagger }}]\sim \lambda
 ^{5},\label{eq33a} \\
 & & [y_{u,d}^{(1)^{\dagger
 }}y_{u,d}^{(1)},y_{u,d}^{(0)^{\dagger }}y_{u,d}^{(0)}]\sim \lambda
 ^{5}.\label{eq33b}
\end{eqnarray}
\label{eq33s}
\end{mathletters}
From Eqs.~(\ref{eq33s}) using the time independent
perturbation theory one obtains that the diagonalizing matrices of the
biunitary transformations in the next order are corrected by the terms
of the order \( \lambda ^{5} \) and this completes the proof.

\section{Evolution of the CKM matrix}

In this section we will find the equations for the one loop evolution
of the squares of the absolute values of the off-diagonal elements of
the CKM matrix.

The CKM matrix is defined in Eq.~(\ref{eq3}). Using the fact that \(
(U_{u})_{L} \) is energy independent we obtain from Eq.~(\ref{eq35})
\begin{equation}
\label{eq42}
(U_{u})_{L}y_{d}(t)y_{d}(t)^{\dagger }(U_{u})^{\dagger }_{L}=
r^{'}_{g}(t)(h(t))^{(2\alpha ^{d}_{3}/\alpha
^{d}_{2})}Z(t)(U_{u})_{L}y_{d} (t_{0})y_{d}(t_{0})^{\dagger
}(U_{u})^{\dagger }_{L}Z(t)
\end{equation}
which can be written in the following way
\begin{equation}
\label{eq43}
V_{CKM}(t)M^{2}_{d}(t)V^{\dagger
}_{CKM}(t)=r^{'}_{g}(t)(h(t))^{(2\alpha ^{d}_{3}/\alpha
^{d}_{2})}Z(t)(U_{u})_{L}y_{d}(t_{0})y_{d}(t_{0})^{\dagger }
(U_{u})_{L}^{\dagger }Z(t)
\end{equation}
where \( M^{2}_{d}(t) \) is the diagonal matrix of the squares of the
physical down quarks Yukawa couplings which become the squares of the
down quark masses after the spontaneous symmetry breaking. Now
differentiating Eq.~(\ref{eq43}) with respect to \( t \) we obtain the
following result
\begin{eqnarray}
&&V_{CKM}^{\dagger}(t)\frac{dV_{CKM}}{dt}\nonumber\\
\label{eq44}
&&= (M^{2}_{d})^{-1}V_{CKM}^{\dagger }(t)
\frac{{dV_{CKM}}}{dt}M^{2}_{d}-(M^{2}_{d})^{-1}V_{CKM}^{\dagger }(t)
\frac{{d}}{dt}((U_{u})_{L}y_{d}(t)y_{d}(t)^{\dagger
}(U_{u})_{L}^{\dagger})
V_{CKM}(t)+(M^{2}_{d})^{-1}\frac{dM^{2}_{d}}{dt}.
\end{eqnarray}
The second term on the right hand side of Eq.~(\ref{eq44}) is equal
\begin{equation}
\label{eq45}
(M^{2}_{d})^{-1}V_{CKM}^{\dagger }(t)
\frac{{d}}{dt}((U_{u})_{L}y_{d}(t)y_{d}(t)^{\dagger}(U_{u})_{L}^{\dagger})
V_{CKM}(t) =\frac{d\ln(h)}{dt}({\bf R}^{\dagger }{\bf R}
+(M_{d}^{2})^{-1}{\bf R}^{\dagger }{\bf R}M_{d}^{2}) +\frac{d\ln
(r^{'}_{g}(t)(h(t))^{(2\alpha ^{d}_{3}/ \alpha ^{d}_{2})})}{dt}I
\end{equation}
where the vector \( {\bf R}=(V_{td},V_{ts},V_{tb}) \) and \( h(t) \)
is given in Eq.~(\ref{eq34}).

The off diagonal matrix elements of \( V_{CKM}^{\dagger
}(t)\frac{dV_{CKM}}{dt} \) can now be evaluated from
Eq.~(\ref{eq44}). Using these matrix elements we obtain the following
evolution equations
\begin{mathletters}
\label{eq37all}
\begin{equation}
\label{37a}
\frac{d|V_{ub}|^{2}}{dt}=-\frac{h^{'}}{h}\frac{m^{2}_{d}+m^{2}_{b}}{m^{2}_{d}-m^{2}_{b}}(V_{ub}^{*}V_{ud}V^{*}_{td}V_{tb})-\frac{h^{'}}{h}\frac{m^{2}_{s}+m^{2}_{b}}{m^{2}_{s}-m^{2}_{b}}(V_{ub}^{*}V_{us}V^{*}_{ts}V_{tb})+\mbox
{c.c}\approx -\frac{2h^{'}}{h}|V_{ub}|^{2}|V_{tb}|^{2},
\end{equation}
\begin{equation}
\label{37b}
\frac{d|V_{cb}|^{2}}{dt}=-\frac{h^{'}}{h}\frac{m^{2}_{d}+m^{2}_{b}}{m^{2}_{d}-m^{2}_{b}}(V_{cb}^{*}V_{cd}V^{*}_{td}V_{tb})-\frac{h^{'}}{h}\frac{m^{2}_{s}+m^{2}_{b}}{m^{2}_{s}-m^{2}_{b}}(V_{cb}^{*}V_{cs}V^{*}_{ts}V_{tb})+\mbox
{c.c}\approx -\frac{2h^{'}}{h}|V_{cb}|^{2}|V_{tb}|^{2},
\end{equation}
\begin{equation}
\label{37c}
\frac{d|V_{td}|^{2}}{dt}=-\frac{h^{'}}{h}\frac{m^{2}_{s}+m^{2}_{d}}{m^{2}_{s}-m^{2}_{d}}(V_{td}^{*}V_{ts}V^{*}_{ts}V_{td})-\frac{h^{'}}{h}\frac{m^{2}_{b}+m^{2}_{d}}{m^{2}_{b}-m^{2}_{d}}(V_{td}^{*}V_{tb}V^{*}_{tb}V_{td})+\mbox
{c.c}\approx -\frac{2h^{'}}{h}|V_{td}|^{2}|(1-|V_{td}|^{2}),
\end{equation}
\begin{equation}
\label{37d}
\frac{d|V_{cd}|^{2}}{dt}=-\frac{h^{'}}{h}\frac{m^{2}_{s}+m^{2}_{d}}{m^{2}_{s}-m^{2}_{d}}(V_{cd}^{*}V_{cs}V^{*}_{ts}V_{td})-\frac{h^{'}}{h}\frac{m^{2}_{b}+m^{2}_{d}}{m^{2}_{b}-m^{2}_{d}}(V_{cd}^{*}V_{cb}V^{*}_{tb}V_{td})+\mbox
{c.c}\approx \frac{2h^{'}}{h}|V_{cd}|^{2}|V_{td}|^{2}.
\end{equation}
\end{mathletters}
Eqs.~(\ref{eq37all}) can be solved explicitly and the
solution reads
\begin{mathletters}
\label{eq38all}
\begin{equation}
\label{eq38a}
|V_{ub}|^{2}=\frac{|V^{0}_{ub}|^{2}}{|V^{0}_{tb}|^{2}(h^2-1)+1},
\end{equation}
\begin{equation}
\label{eq38b}
|V_{cb}|^{2}=\frac{|V^{0}_{cb}|^{2}}{|V^{0}_{tb}|^{2}(h^2-1)+1},
\end{equation}
\begin{equation}
\label{eq38c}
|V_{td}|^{2}=\frac{|V^{0}_{td}|^{2}}{|V^{0}_{td}|^{2}(1-h^2)+h^2},
\end{equation}
\begin{equation}
\label{eq38d}
|V_{cd}|^{2}=\frac{|V^{0}_{cd}|^{2}}{|V^{0}_{td}|^{2}(1-h^2)+h^2}.
\end{equation}
\end{mathletters}
Here \( |V^{0}_{ij}|^{2} \) are the initial values of the squares of
the absolute values of the corresponding CKM matrix elements and \( h
\) is given in Eq.~(\ref{eq34}).  From the unitarity of the CKM matrix
and using~Eqs.~(\ref{eq38all}) one can calculate the absolute values
of all the remaining matrix elements of the CKM matrix and thus we
determine the renormalization group evolution of the full CKM matrix.

\section{Conclusions}

In this paper we have analyzed the renormalization group evolution of
the quark Yukawa couplings and the CKM matrix based on the observed
hierarchy of the quark masses and the CKM matrix. The inclusion of the
hierarchy greatly simplifies the analysis and leads to simple,
explicit results for the evolution of the Yukawa couplings
(Eqs.~(\ref{eq20}) and Eqs.~(\ref{eq35})) and the CKM matrix
(Eqs.~(\ref{eq38all})). The other remarkable result is that the
diagonalizing matrices of the up quark Yukawa couplings are energy
independent in the leading order. This means that the transformation
\( (\psi _{u})_{L,R}\rightarrow (U_{u})_{L,R}(\psi _{u})_{L,R} \), \(
(\psi _{d})_{L,R}\rightarrow (U_{u})_{L,R}(\psi _{d})_{L,R} \) will
diagonalize the matrix of the up quark Yukawa couplings and it will
stay diagonal upon the renormalization group evolution and the CKM
matrix will be determined only from the down quarks Yukawa
couplings. This fact may simplify the model building based on the
symmetries of the quark Yukawa couplings.

\acknowledgments This paper is partially supported by the CoNaCyT
(Mexico) project 3512P-E9608.  S.R.J.W. gratefully acknowledges
partial support by Comisi\'{o}n de Operaci\'{o}n y Fomento de
Actividades Acad\'{e}micas - COFAA (Instituto Polit\'{e}cnico
Nacional). J.G.~Mora~H.\ thanks dr.~A.~Odzijewicz for the hospitality
in the Institute of Theoretical Physics of the Bia{\l}ystok University
(Poland) during the preparation of the paper.

\appendix
\section*{Constants in the renormalization group equations}

The two loop renormalization group equations for various models have
the following structure
\begin{equation}
\frac{dg_l}{dt}=\frac{1}{(4\pi)^2}b_lg_l^3-\frac{1}{(4\pi)^4}G_lg_l^3,
\end{equation}
\begin{equation}
\frac{dy_{u,d,e,\nu}}{dt}=
\left[\frac{1}{(4\pi)^2}\beta_{u,d,e,\nu}^{(1)}
+\frac{1}{(4\pi)^4}\beta_{u,d,e,\nu}^{(2)} \right]y_{u,d,e,\nu}
\end{equation}
where $b_l$ are constants dependent on the model and $G_l$,
$\beta_{u,d,e,\nu}^{(1)}$ and $\beta_{u,d,e,\nu}^{(2)}$ are the
following functions of the coupling constants and the squares of the
Yukawa couplings
$H^{(1)}_{u,d,e,\nu}=y_{u,d,e,\nu}y^{\dagger}_{u,d,e,\nu}$
\begin{equation}
G_{l}=C_{lu}\mbox{Tr}(H_{u}^{(1)})+\sum_{k}b_{kl}g_{k}^{2},
\end{equation}
\begin{eqnarray}
\beta^{(1)}_l&=&\alpha_1^{l}(t) +\alpha_2^{l}H_u^{(1)}
+\alpha_3^{l}\mbox{Tr}(H_u^{(1)}) +\alpha_4^{l}H_d^{(1)}
+\alpha_5^{l}\mbox{Tr}(H_d^{(1)})\nonumber\\
 &+&\alpha_6^{l}H_e^{(1)}
+\alpha_7^{l}\mbox{Tr}(H_e^{(1)}) +\alpha_8^{l}H_\nu^{(1)}
+\alpha_9^{l}\mbox{Tr}(H_\nu^{(1)}),
\end{eqnarray}
\begin{eqnarray}
\beta^{(2)}_l&=&\sum_{lkn}B_{lkn}^1g_k^2g_n^2
+\sum_{lkn}B_{lkn}^2g_k^2H_n^{(1)}
+\sum_{lkn}B_{lkn}^3g_k^2\mbox{Tr}(H_n^{(1)})\nonumber\\
&+&\sum_{lkn}B_{lkn}^4 H_k^{(1)} H_n^{(1)} +\sum_{lkn}B_{lkn}^5
H_k^{(1)} \mbox{Tr}(H_n^{(1)}).
\end{eqnarray}
The functions $\alpha_1^{l}(t)$ are equal
\begin{enumerate}
\item[A.] for the SM and DHM
\begin{eqnarray}
&&\alpha_1^u(t)=-(\frac{17}{20}g_1^2+\frac{9}{4}g_2^2+8g_3^2),\\
&&\alpha_1^d(t)=-(\frac{1}{4}g_1^2+\frac{9}{4}g_2^2+8g_3^2),\\
&&\alpha_1^e(t)=-(\frac{9}{20}g_1^2+\frac{9}{4}g_2^2),\\
&&\alpha_1^\nu(t)=-(\frac{9}{20}g_1^2+\frac{9}{4}g_2^2).
\end{eqnarray}
\item[B.] for the MSSM
\begin{eqnarray}
&&\alpha_1^u(t)=-(\frac{13}{15}g_1^2+3g_2^2+\frac{16}{3}g_3^2),\\
&&\alpha_1^d(t)=-(\frac{7}{15}g_1^2+3g_2^2+\frac{16}{3}g_3^2),\\
&&\alpha_1^e(t)=-(\frac{9}{5}g_1^2+3g_2^2),\\
&&\alpha_1^\nu(t)=-(\frac{3}{5}g_1^2+3g_2^2).
\end{eqnarray}
\end{enumerate}
The values of the coefficients $b_l$ and $\alpha_k^l$ for the one loop
renormalization group equations are given in Tables~\ref{table1},
\ref{table2} and~\ref{table3} (see e.g.~\cite{ref1,ref2,ref3}).
\narrowtext
\begin{table}[h]
\caption{Coefficients $b_l$ for various models.}
\begin{tabular}{lccc}
&$b_1$&$b_2$&$b_3$\\ 
\hline 
SM&$\frac{41}{10}$&$-\frac{19}{6}$&$-7$\\
DHM&$\frac{21}{5}$&$-3$&$-7$\\
MSSM&$\frac{33}{5}$&$1$&$-3$
\label{table1}
\end{tabular}
\end{table}
\begin{table}[h]
\caption{The coefficients $\alpha_k^l$ for various models. The model
dependent constants $a$, $b$ and $c$ are given in Table~\ref{table3}.}
\begin{tabular}{lcccccccc}
$l$& $\alpha_2^l$& $\alpha_3^l$& $\alpha_4^l$& $\alpha_5^l$&
$\alpha_6^l$& $\alpha_7^l$& $\alpha_8^l$& $\alpha_9^l$\\
\hline
$u$&$\frac{3}{2}b$& $3$& $\frac{3}{2}c$& $3a$& $0$& $a$& $0$& $1$\\
$d$&$\frac{3}{2}c$& $3a$& $\frac{3}{2}b$& $3$& $0$& $1$& $0$& $a$\\
$e$&$0$& $3a$& $0$& $3$& $\frac{3}{2}b$& $1$& $\frac{3}{2}c$& $a$\\
$\nu$&$0$& $3$& $0$& $3a$& $\frac{3}{2}c$& $a$& $\frac{3}{2}b$& $1$
\end{tabular}
\label{table2}
\end{table}
\begin{table}[!]
\caption{The constants $a$, $b$ and $c$ for various models.}
\begin{tabular}{lccc}
Model&$a$&$b$&$c$\\
\hline
SM&$1$&$1$&$-1$\\
DHM&$0$&$1$&$\frac{1}{3}$\\
MSSM&$0$&$2$&$\frac{2}{3}$
\end{tabular}
\label{table3}
\end{table}
\widetext

The coefficients for the two loop renormalization group equations 
that appear in the functions $G_{l}$ and $\beta^{(2)}_{l}$ can
be found in Ref.~\cite{ref7}.

\end{document}